\documentclass[12pt,english,preprint]{aastex} 
\usepackage{natbib}
\begin{document}

\title{Halting planet migration by photoevaporation from the central
source}

\author{Isamu Matsuyama\altaffilmark{1}, Doug Johnstone\altaffilmark{2},
	\and Norman Murray\altaffilmark{3,4}}

\altaffiltext{1}{Department of Astronomy and Astrophysics,University of Toronto,
Toronto, ON M5S 3H8, Canada; isamu@astro.utoronto.ca}
\altaffiltext{2}{National Research Council Canada, Herzberg Institute of
Astrophysics, 5071 West Saanich Road, Victoria, BC V9E 2E7, Canada; doug.johnstone@nrc.ca}
\altaffiltext{3}{Canadian Institute for Theoretical Astrophysics, University of
Toronto, Toronto, ON M5S 3H8, Canada; murray@cita.utoronto.ca}
\altaffiltext{4}{Canada Research Chair in Astrophysics}

\begin{abstract}
The recent discovery of Jupiter-mass planets orbiting at a few AU from
their stars compliments earlier detections of massive planets on very
small orbits.  The short period orbits strongly suggest that planet
migration has occurred, with the likely mechanism being tidal
interactions between the planets and the gas disks out of which they
formed.  The newly discovered long period planets, together with the
gas giant planets in our solar system, show that migration is either
absent or rapidly halted in at least some systems.  We propose a
mechanism for halting type-II migration at several AU in a gas disk.  
Photoevaporation of the disk by irradiation from the central star can produce a gap in the
disk at a few AU, preventing planets outside the gap from migrating down to the
star.  This would result in an excess of systems with planets at or
just outside the photoevaporation radius. 
\end{abstract}


\section{Introduction}

It is believed that gas giants do not generally form at small orbital
distances from the central star \citep{1995Sci...267..360B}.  Thus, a
natural explanation for extra solar planets orbiting close to the
central star is that these planets formed further away in the
protoplanetary disk and migrated inward to where they are now
observed.  A variety of mechanisms have been proposed to explain
planet migration: the interaction between a planet and a planetesimal
disk \citep{1998Sci...279...69M}, the gravitational interaction
between two or more Jupiter mass planets \citep{1996Sci...274..954R},
and the tidal gravitational interaction between the planet and the
surrounding disk gas \citep{1979ApJ...233..857G, 1980ApJ...241..425G}.
The last mechanism, focused on in this paper, is expected to be
dominant at early times since the surrounding gaseous disk is required
for the formation of planets.  

If the perturbation exerted on the disk by the  planet is small, the disk structure 
is not greatly altered, and the planet moves inward relative to the 
surrounding gas \citep{1997Icar..126..261W}.
This type of migration is referred to as `type-I'.  
However, if the planet is large it may open a gap in the disk \citep{1980ApJ...241..425G}.  
The planet is locked to the disk and moves either inward or outward in lock-step with the 
gaseous disk.
This slower migration is referred to as `type-II'.

We propose a mechanism for halting type-II migration: 
photoevaporation driven by radiation from the central star.  
The planet's final location is consistent with the Solar System
and the growing class of extra solar planets with nearly circular orbits outside 
of a few AU \citep[see][figure 4]{2002ApJ...571..528T}.
Photoevaporation by the central star was proposed by
\citet{1993Icar..106...92S} and \citet{1994ApJ...428..654H} as a way to remove a gas disk.
\citet{2000prpl.conf..401H} generalized the discussion, describing the variety 
of possible disk removal mechanisms:  accretion, planet formation, stellar encounters, 
stellar winds or disk winds, and photoevaporation by ultraviolet photons from the central source or 
massive external stars.
\citet{2000prpl.conf..401H} concluded that the dominant mechanisms for a wide
range of disk sizes are viscous accretion and photoevaporation,
operating in concert within the disk.  
In this paper, we consider photoevaporation by the central source and viscous accretion.

\section{Model}

The model for disk removal used here is developed in a paper by
\citet*{disk.photoevaporation} and is similar to that used by
\citet*{2001MNRAS.328..485C}.  Additionally, we assume a planet with a large mass,
which opens a narrow gap in the disk, and assume that planet migration
proceeds in lock-step with the disk evolution (type-II migration).  
The gas disk orbiting the central star with the local Keplerian circular velocity, $v_k$, 
is axisymmetric and geometrically thin.
Considering angular momentum and mass conservation of a disk annulus at a radius, 
$ R $, with kinematic viscosity, $ \nu $, the disk surface density evolution can be
described by \citep{1981ARA&A..19..137P}
\begin{equation}\label{sigma evolution} 
\frac{\partial \Sigma }{\partial t}=\frac{3}{R}\frac{\partial}
{\partial R}\left[ R^{1/2}\frac{\partial }{\partial R}
\left( \nu \Sigma R^{1/2}\right) \right],
\end{equation} 
where $ \Sigma $ is the disk surface density and $ t $ is the disk
evolutionary time.  We adopt the standard $ \alpha $-parameterization
of \citet{1973A&A....24..337S} and write $ \nu=\alpha c_{s}H, $ 
where $ c_{s} $ is the sound speed at the disk mid-plane and $ H $ is the disk thickness.  
\citet{1998ApJ...495..385H} estimate $ \alpha \sim 10^{-2}-10^{-3} $ and 
a disk mass $\sim0.01-0.2M_{\sun } $ for T Tauri stars (TTSs).  
For the modeled disk temperature distribution, $T_{d}\propto R^{-1/2} $ 
\citep{1998ApJ...500..411D}, the viscosity takes the simple form, $ \nu (R)\propto R $. 
Given a solution for equation \ref{sigma evolution}, we find the drift velocity,
\begin{equation}\label{drift velocity} 
v_{R}=-\frac{3}{\Sigma R^{1/2}}\frac{\partial }{\partial R}(\nu \Sigma
R^{1/2}),
\end{equation} 
and describe the evolution of the disk stream lines. 

The EUV ($ \lambda<912$\AA ) photons from the central star and the
accretion shock (we will refer to these photons as the EUV photons
from the central source) are capable of ionizing hydrogen and
evaporating material from the disk surface.  
Photoevaporation forms an ionized atmosphere above the thin viscous disk.  
\citet{1994ApJ...428..654H} found analytic solutions for the photoevaporation mass 
loss rate by EUV photons from the central source.  
These photons are attenuated by recombined hydrogen atoms and scattered in the 
ionized atmosphere, providing a source of diffuse EUV photons that dominates 
the flux onto the disk at radii much larger than the size of the central star.  
Disk material is gravitationally bound to the central star inside the gravitational
radius, $ R_{g}=GM_\star/3c_s'^{2} $, and it flows out of the disk 
surface at the sound speed of the heated material, $ c_s' $, beyond this radius.
Given the number density of ionized hydrogen at the base of the ionized layer, $ n(R)
$, we can calculate the evaporation rate:
\begin{equation}\label{central dsigmadt} 
\dot{\Sigma}_{ph}(R)=\cases{2\mu m_pn(R)c_s', &if $R>R_g$;\cr 0, &otherwise, \cr}
\end{equation} 
where $ \mu=0.68 $ is the mean molecular weight for the ionized
material.

Assuming ionization equilibrium \citet{1994ApJ...428..654H} found the
number density at the base of the ionized layer, for $ R>R_g $:
\begin{equation}\label{number density} 
n(R)=3.1\times 10^{5}\left( \frac{\phi
}{10^{40}\textrm{s}^{-1}}\right) ^{1/2}\left(
\frac{R_g}{1AU}\right) \left( \frac{R}{1AU}\right)
^{-5/2}\textrm{cm}^{-3};
\end{equation} 
where $ \phi $ is the total ionizing flux from the central source.  We
assume that half of the accretion luminosity is radiated as hot
continuum from accretion shocks and that the EUV flux can be
characterized as blackbody emission at $ T_{as}=10^{4} $K
\citep{2000ApJ...544..927G, 2000ApJ...539..815J, 1989ApJ...344..925K}.
There are currently no strong observational constraints on the
ionizing flux from the central star, $\phi_{\star}$.  The EUV flux for the
sun is $\sim2.6\times 10^{37}\textrm{s}^{-1} $ \citep[see][figure
10b]{1998A&A...334..685W}. Since TTSs are chromospherically active
\citep{1998ApJ...500..411D}, their EUV fluxes must be higher by few
orders of magnitude. We estimate the ionizing flux from the FUV
spectra of TTSs \cite[figure 3]{2000ApJS..129..399V} and the
corresponding distances to the star \cite[table
1]{2000ApJ...539..815J}.  
The typical values are $\sim10^{39}\textrm{s}^{-1} $ for classical T Tauri stars. 
Extrapolating the FUV flux underestimates the EUV flux; 
thus the EUV flux might be as high as $10^{40}\textrm{s}^{-1}$.

Combining photoevaporation with viscous accretion is done numerically.
At each time step, photoevaporation induced mass loss and viscous
diffusion induced disk evolution are solved. 
For details on the disk dispersal calculation refer to \citet{disk.photoevaporation}.
The simulation is stopped when the disk mass is $1M_J$. 

\section{Discussion}

\begin{figure}
\plotone{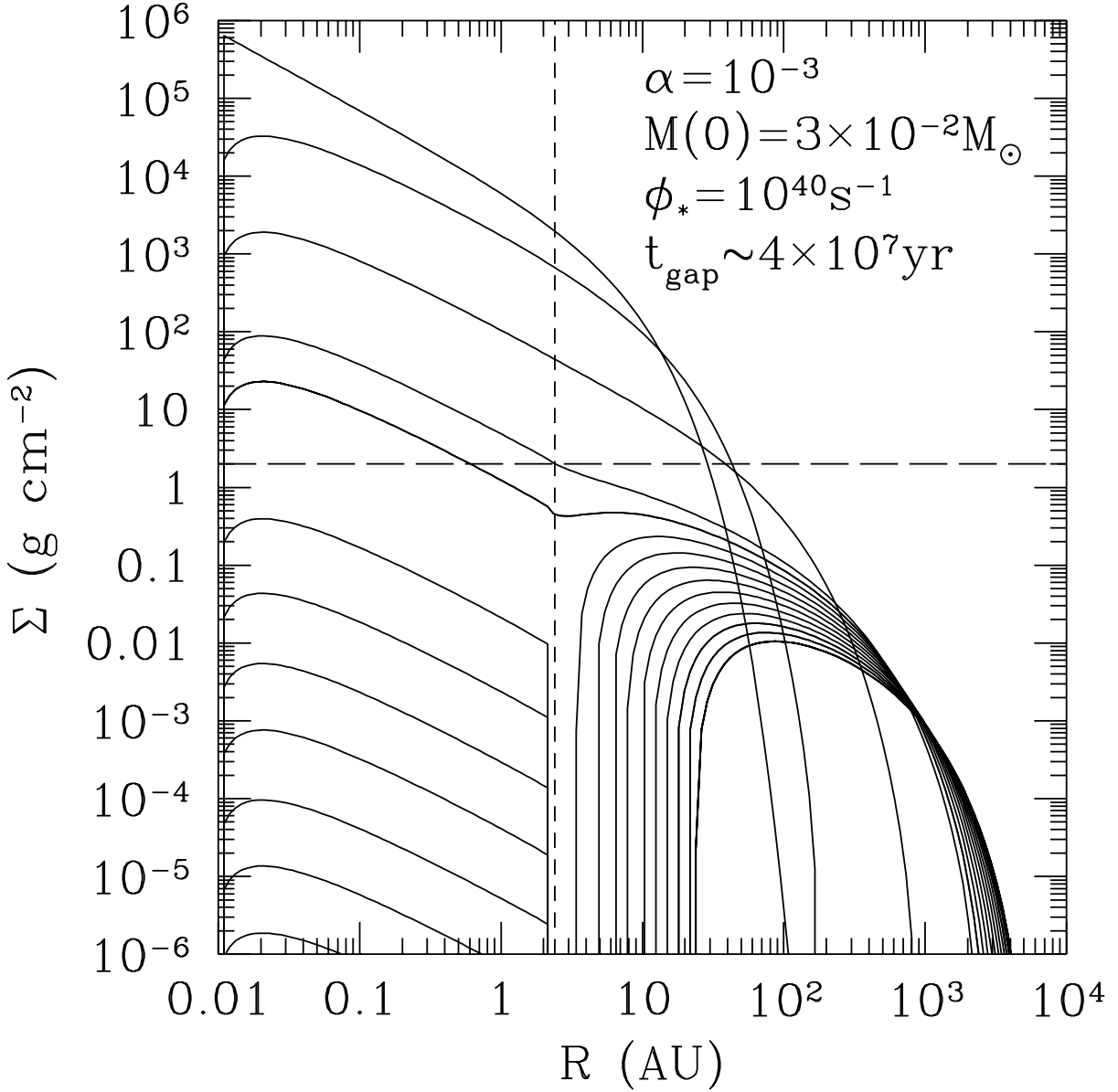}
\caption{Snapshots of the surface density for a fiducial model under
the influence of viscous diffusion and photoevaporation from
the central source.  The model corresponds to \protect\protect$
\alpha=10^{-3}\protect \protect $, and an initial disk mass,
\protect\protect$ M(0)=0.03M_{\odot }\protect \protect $.  The
short-dashed line indicates the location of the gravitational radius
and the long-dashed line corresponds to the minimum surface density
for gap formation by photoevaporation.  The solid curves represent
\protect\protect$ t=0,10^{6},10^{7},3.6\times10^{7}, 4.2\times
10^{7},4.7\times 10^{7},5.2\times10^{7},5.7\times10^{7},6.2\times
10^{7}, 6.7\times 10^{7},7.2\times 10^{7},7.7\times
10^{7},8.2\times10^{7},8.7\times10^{7}, \textrm{and} 9.2\times
10^{7}\textrm{yr}\protect \protect $.  The gap structure starts
forming at \protect\protect$ t_{gap}\sim 4\times 10^{7}\protect
\protect $yr, when the disk mass is \protect\protect$ \sim 3\times
M_J\protect \protect $ and the surface density at the gravitational radius
is \protect\protect$ \Sigma_{gap}\sim2\textrm{g\,cm}^{-2}\protect \protect $ 
(long-dashed line).
\label{sigma}}
\end{figure}

At a given disk radius, a gap forms when the mass transported by
viscous accretion is equal to the mass removed by photoevaporation.
This occurs when the photoevaporation time scale, \(
\Sigma/\dot{\Sigma }_{ph} \), is equal to the viscous diffusion time
scale, \( R^{2}/3\nu \). Therefore, we can estimate the
surface density for gap formation at the gravitational radius,
\begin{equation}\label{sigma gap} 
\Sigma _{gap}\sim(2\textrm{\,g\,cm}^{-2})\left( \frac{\alpha }{10^{-3}}\right) ^{-1}\left( \frac{\phi _{\star }}{10^{40}\textrm{s}^{-1}}\right)^{1/2}
\left( \frac{R_{g}}{2.4\textrm{AU}}\right) ^{-1},
\end{equation} 
Figure \ref{sigma} shows snapshots of the disk surface density
distribution for a disk with initial mass, $M(0)=0.03M_{\odot }$ and $
\alpha =10^{-3} $.  A gap starts forming at the gravitational radius
when the surface density reaches $\Sigma_{gap}\sim2\textrm{\,g\,cm}^{-2}$
during the last stages ($ t_{gap}\sim4\times 10^{7}
$yr) of the disk evolution and the disk is divided into two annuli.
The subsequent evolution is dominated by two competing effects; 
viscous diffusion attempts to spread both annuli and remove the gap structure 
while photoevaporation removes material predominantly at the gravitational radius, 
reopening the gap.  The material in the inner annulus is quickly removed both at the inner
edge, by accretion onto the central star, and at the outer edge, by
the combination of viscous spreading and photoevaporation.  In
contrast, the outer annulus loses material primarily from its inner
edge. The inner edge of the outer annulus recedes from the star as
photoevaporation removes disk material, thereby reducing the mass
removal rate in accordance with equations (\ref{central dsigmadt}) and
(\ref{number density}).

\begin{figure}
\plotone{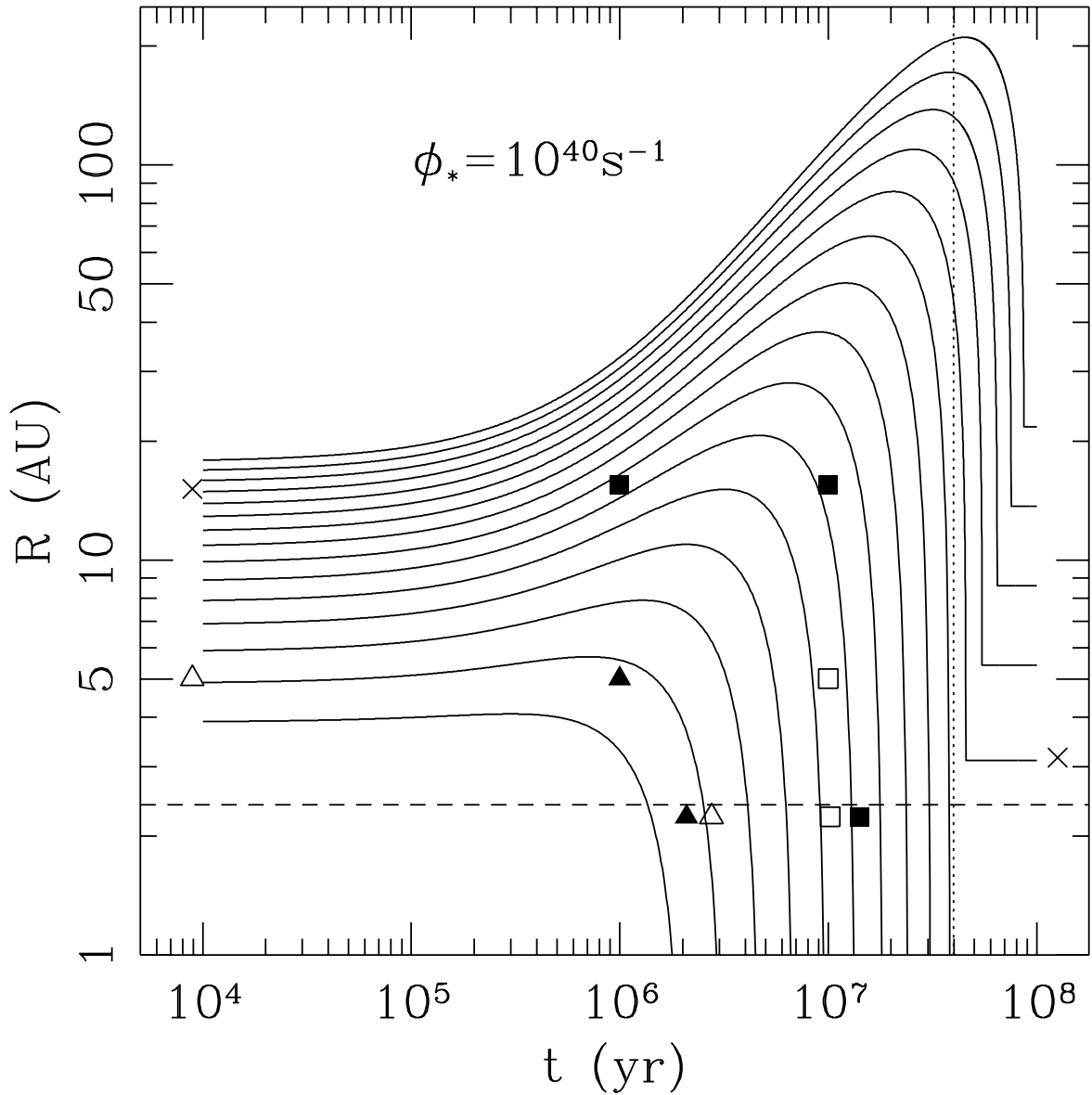}
\caption{Disk stream lines for the fiducial model with
\protect\protect$ M(0)=0.03M_{\odot }\protect \protect $ and
\protect\protect$ \alpha =10^{-3}\protect \protect $.  The initial
spacing (at \protect\protect$ t=10^{4}\protect \protect $yr) between
the streamlines is \protect\protect$ 1\protect \protect $AU,
covering the disk from 4AU to 18AU. The short-dashed line
indicates the location of the gravitational radius,
\protect\protect$R_g=2.4\protect \protect $AU, and the dotted line
represents the time when the photoevaporation gap starts forming,
\protect\protect$ t_{gap}\sim4\times 10^{7}\protect \protect $yr.
Planets forming at $R\sim 5\,$AU and tidally locking to the disk early
($t =10^4\,$yr) (open triangles) or late (solid triangles) fall onto
the central star at $t \sim 2.5\times 10^6\,$yr.  In contrast, a
planet starting with initial semi-major axis $R\sim 15$AU at $t =
10^4\,$yr (crosses) stops migrating when it reaches the
photoevaporation gap.
\label{radius.1e4}}
\end{figure}

\begin{figure}
\plotone{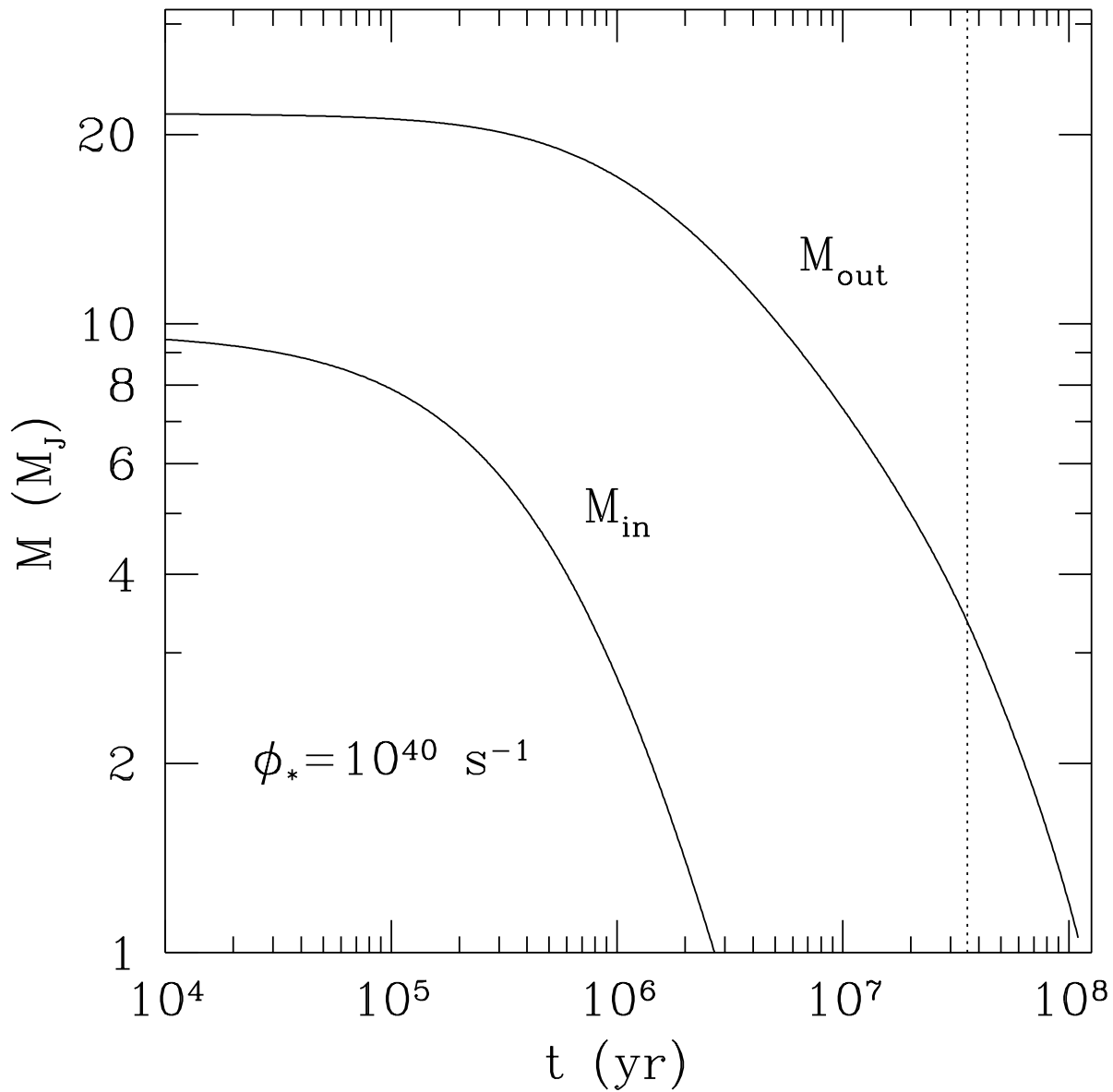}
\caption{Disk mass outside the gravitational radius, \protect\protect$
M_{out}\protect \protect $, and disk mass inside the gravitational
radius, \protect\protect$M_{in}\protect \protect $, as a function of
disk lifetime for the fiducial model.  The dotted line indicates the
time, \protect\protect$ t_{gap}\sim4\times10^{7}\protect \protect $yr, 
when the gap structure starts forming.
\label{mass}}
\end{figure}

We model a planet which has opened a narrow gap in the
disk. The migrating planet will open a gap if its mass is greater than
the critical mass,
$ M_c\sim M_\star\left(81\pi\alpha c_s^2/8v_k^2\right) $, 
\citep[see, e.g., ][]{1993prpl.conf..749L}.
The planet is initially tidally locked to the disk and moves with the
gas, migrating outward or inward depending on its initial location.
This migration may stop for two reasons.  First, the combination of
photoevaporation from the central source and viscous spreading removes
disk mass, reducing the gravitational interaction between the planet
and the disk and allowing the planet to decouple its orbit from that
of the gas.  Second, the planet will stop if it reaches the gap
created by photoevaporation, as the tidal torques are unable to
influence the planet across such a large opening.  In order to
describe planet migration, we show in figure \ref{radius.1e4} the disk
stream lines which start between $ 4 $ and $ 18 $AU, at $ t=10^{4}
$yr, where there is enough mass to form a Jupiter mass planet in an
annulus of width $ 4 $AU.  After the planet opens a gap it migrates
along the disk stream lines; therefore, we can follow planet migration
in figure \ref{radius.1e4} for different planet gap opening locations
and times.  The disk mass outside the gravitational radius is reduced
to Jupiter's mass in $\sim 10^{8} $yr (see figure \ref{mass}); therefore,
a migrating planet with Jupiter's mass stops migrating at any disk
location soon after this time.  However, a migrating planet may reach the
inner disk boundary ($ R=10^{-2} $AU) or the photoevaporation gap
before the disk mass is reduced to Jupiter's mass.

Figure \ref{radius.1e4} provides information on which initial locations, 
and times, allow a planet tidally locked to the disk to survive. 
It is difficult to time the gaseous disk removal in such a manner as
to maintain planets at large distances from the central star.  
A planet which opens a narrow gap at $ t=10^{4} $yr only survives if its 
initial disk location is greater than the critical radius $\sim14$ AU, 
because the planet reaches the outer edge of the photoevaporation gap 
after the gap has formed.
Since the disk expands rapidly at early times, this critical radius increases 
with later planet gap opening timescales; i.e. the later the planet forms, 
the further away from the central star it has to be in order to survive.
However, this behavior changes after the photoevaporation induced gap forms.  
If the planet forms at a time similar to when the photoevaporation gap forms, 
$t_{gap}\sim4\times 10^{7} $, it survives as long as it originates outside 
the gravitational radius.

\section{Conclusions}

In this letter, we have studied different scenarios in which a
planet undergoing type-II migration survives due to the formation of a gap by
photoevaporation from the central source.  
If the planet gap opening occurs close to when the photoevaporation gap forms (at $ t\sim
4\times 10^{7}\,$yr for our fiducial model), the final semi-major axis
of the planet is near the gravitational radius ($ \sim 2.4 $AU) for a
wide range of initial semi-major axes.   
This scenario requires long time scales for planet formation 
($ 10^{6}-10^{8} $yr), similar to those predicted by the core accretion model
\citep{1996Icar..124...62P,2000Icar..143....2B}.

We can also predict the final semi-major axis when considering rapid planet 
formation by disk instability \citep{2000ApJ...536L.101B}.  
Our disk initially spreads outward under the influence of viscosity, allowing planets 
that form early at small radii to move out beyond $R_g$ before the photoevaporation gap 
opens, and enhancing their survival rate. 
The rate and amount of viscous spreading in real disks depends on the initial surface 
density profile and any continuing accretion onto the disk.
If the planet induced gap opening occurs at early stages of the disk evolution 
(for example, at $ t\sim 10^{4}$yr), all the planets initially located inside 
$ \sim 14 $AU will reach the disk inner boundary at the central star.  
The only planets which survive are those which form at a narrow range of semi-major
axes ($ 14\la R\la 18 $AU).  
However, viscous diffusion spreads this narrow range during planet migration, 
and the final semi-major axis distribution is $ 3\la R\la 22 $AU.  
This final distribution of planets has similarities to our own Solar System and it
suggests that the gas giants could have formed at $\sim 14-18$ AU and migrated to 
their current locations.  
Furthermore, the initial planet locations are in agreement with recent simulations 
of planet formation by disk fragmentation \citep{2002Sci...298.1756M}. 

The minimum surface density for gap opening by photoevaporation, $\Sigma_{gap}$,
and the disk size, $R_d$, determine the disk mass outside the gravitational radius.  
In our fiducial model, the mass outside the gravitational radius at the time of gap 
opening is $2\pi\Sigma_{gap}R_g(R_d-R_g)\sim3M_J $; therefore it is possible to 
consider migration of a Jupiter mass planet.  
However, for lower ionizing fluxes (i.e. $\phi_{\star}\la 10^{40}\textrm{s}^{-1}$), 
the minimum surface density for gap opening by photoevaporation becomes smaller (see 
equation \ref{sigma gap}) and the halting of planet migration becomes marginal.  
For example, for $\phi_{\star}=10^{38}\textrm{s}^{-1}$, the minimum surface density for 
gap opening and the disk mass outside the gravitational radius are smaller by one order 
of magnitude ($\Sigma_{gap}\sim 0.2\textrm{gcm}^{-2}$ and $M_{out}=0.3M_J$).

We note that the gravitational radius, $R_g$, roughly coincides with the
location of the asteroid belt in the solar system. It is well known
that the surface density of heavy elements reaches a severe local
minimum in the asteroid belt; this is usually attributed to long term
gravitational effects from Jupiter, although it is not clear that the
amount of depletion that results is sufficient  
\citep{1997Sci...275..375L, 1996AJ....112.1278H}.
Simple estimates show that the planetesimal formation time scale is much shorter than the
gap formation time, so it is not clear that the formation of a
photoevaporation gap would significantly reduce the surface density of
planetesimals, but the coincidence is intriguing.

The photoevaporation gap opening timescale ($\sim4\times10^7$yr) is long compared 
to the typically observed  gaseous disk lifetimes ($10^{6}-10^{7}$yr). 
However, the disk lifetime for stars outside stellar clusters is considerably 
longer than the disk lifetime for stars in the hostile stellar clusters 
\citep{disk.photoevaporation}. 
In addition, there is evidence that some stars retain disks between $10^{7}-10^{8}$yr 
\citep[see][figure 2]{Hillenbrand:2002ad}.

It is also possible to halt both type-I and type-II migration in stellar clusters 
by removing the disk. This is a likely situation for the disks surrounded by ionization fronts in 
the Trapezium Cluster \citep{1998ApJ...499..758J, 1999ApJ...515..669S, 2000AJ....119.2919B}.
\citet{1999ApJ...515..669S}, \citet{2000prpl.conf..401H} and \citet{disk.photoevaporation} show
that the disk lifetime due to photoevaporation by external stars is 
in the range $ 10^{5}-10^{6}\,\textrm{yr} $ in the neighborhood of massive O stars, 
and between $10^{6}-10^{7}\textrm{yr} $ at reasonable distances  ($ \sim0.03 - 0.3 $pc)
from external O stars.  
These timescales are in agreement with observational estimates for the lifetime of 
protoplanetary disks in stellar clusters \citep{2001ApJ...553L.153H}.  
Since the outer disk is removed first, there is no photoevaporation gap formation.
This scenario is possible when considering the rapid planet formation by disk instability; 
on the other hand, it is uncertain whether gas giant planets form by core accretion 
in these short lived disks. 

\acknowledgements{}

The research of I.M. is supported by the University of Toronto fellowship and the international 
recruitment award. 
The research of D.J. is partially supported through an NSERC grant held at the University of 
Victoria.
The research of N.M. is supported by the Canada Research Chair Program and by NSERC. 

\bibliography{planet.formation,planet.migration,observations,accretion,numerical,evaporation,observations.planets.fewAU,solar.system}

\begin{thebibliography}{41}
\expandafter\ifx\csname natexlab\endcsname\relax\def\natexlab#1{#1}\fi

\bibitem[{{Bally} {et~al.}(2000){Bally}, {O'Dell}, \&
  {McCaughrean}}]{2000AJ....119.2919B}
{Bally}, J., {O'Dell}, C.~R., \& {McCaughrean}, M.~J. 2000, \aj, 119, 2919

\bibitem[{{Bodenheimer} {et~al.}(2000){Bodenheimer}, {Hubickyj}, \&
  {Lissauer}}]{2000Icar..143....2B}
{Bodenheimer}, P., {Hubickyj}, O., \& {Lissauer}, J.~J. 2000, Icarus, 143, 2

\bibitem[{{Boss}(1995)}]{1995Sci...267..360B}
{Boss}, A.~P. 1995, Science, 267, 360

\bibitem[{{Boss}(2000)}]{2000ApJ...536L.101B}
---. 2000, \apjl, 536, L101

\bibitem[{{Clarke} {et~al.}(2001){Clarke}, {Gendrin}, \&
  {Sotomayor}}]{2001MNRAS.328..485C}
{Clarke}, C.~J., {Gendrin}, A., \& {Sotomayor}, M. 2001, \mnras, 328, 485

\bibitem[{{D'Alessio} {et~al.}(1998){D'Alessio}, {Canto}, {Calvet}, \&
  {Lizano}}]{1998ApJ...500..411D}
{D'Alessio}, P., {Canto}, J., {Calvet}, N., \& {Lizano}, S. 1998, \apj, 500,
  411

\bibitem[{{Goldreich} \& {Tremaine}(1979)}]{1979ApJ...233..857G}
{Goldreich}, P., \& {Tremaine}, S. 1979, \apj, 233, 857

\bibitem[{{Goldreich} \& {Tremaine}(1980)}]{1980ApJ...241..425G}
---. 1980, \apj, 241, 425

\bibitem[{{Gullbring} {et~al.}(2000){Gullbring}, {Calvet}, {Muzerolle}, \&
  {Hartmann}}]{2000ApJ...544..927G}
{Gullbring}, E., {Calvet}, N., {Muzerolle}, J., \& {Hartmann}, L. 2000, \apj,
  544, 927

\bibitem[{{Haisch} {et~al.}(2001){Haisch}, {Lada}, \&
  {Lada}}]{2001ApJ...553L.153H}
{Haisch}, K.~E., {Lada}, E.~A., \& {Lada}, C.~J. 2001, \apjl, 553, L153

\bibitem[{{Hartmann} {et~al.}(1998){Hartmann}, {Calvet}, {Gullbring}, \&
  {D'Alessio}}]{1998ApJ...495..385H}
{Hartmann}, L., {Calvet}, N., {Gullbring}, E., \& {D'Alessio}, P. 1998, \apj,
  495, 385

\bibitem[{Hillenbrand(2002)}]{Hillenbrand:2002ad}
Hillenbrand, L.~A. 2002, astro-ph/0210520

\bibitem[{{Hollenbach} {et~al.}(1994){Hollenbach}, {Johnstone}, {Lizano}, \&
  {Shu}}]{1994ApJ...428..654H}
{Hollenbach}, D., {Johnstone}, D., {Lizano}, S., \& {Shu}, F. 1994, \apj, 428,
  654

\bibitem[{{Hollenbach} {et~al.}(2000){Hollenbach}, {Yorke}, \&
  {Johnstone}}]{2000prpl.conf..401H}
{Hollenbach}, D.~J., {Yorke}, H.~W., \& {Johnstone}, D. 2000, Protostars and
  Planets IV, 401

\bibitem[{{Holman} \& {Murray}(1996)}]{1996AJ....112.1278H}
{Holman}, M.~J., \& {Murray}, N.~W. 1996, \aj, 112, 1278

\bibitem[{{Johns-Krull} {et~al.}(2000){Johns-Krull}, {Valenti}, \&
  {Linsky}}]{2000ApJ...539..815J}
{Johns-Krull}, C.~M., {Valenti}, J.~A., \& {Linsky}, J.~L. 2000, \apj, 539, 815

\bibitem[{{Johnstone} {et~al.}(1998){Johnstone}, {Hollenbach}, \&
  {Bally}}]{1998ApJ...499..758J}
{Johnstone}, D., {Hollenbach}, D., \& {Bally}, J. 1998, \apj, 499, 758

\bibitem[{{Kenyon} {et~al.}(1989){Kenyon}, {Hartmann}, {Imhoff}, \&
  {Cassatella}}]{1989ApJ...344..925K}
{Kenyon}, S.~J., {Hartmann}, L., {Imhoff}, C.~L., \& {Cassatella}, A. 1989,
  \apj, 344, 925

\bibitem[{{Lin} \& {Papaloizou}(1993)}]{1993prpl.conf..749L}
{Lin}, D.~N.~C., \& {Papaloizou}, J.~C.~B. 1993, in Protostars and Planets III,
  749--835

\bibitem[{{Liou} \& {Malhotra}(1997)}]{1997Sci...275..375L}
{Liou}, J.~C., \& {Malhotra}, R. 1997, Science, vol.~275, p.~375-377 (1997).,
  275, 375

\bibitem[{{Matsuyama} {et~al.}(2003){Matsuyama}, {Johnstone}, \&
  {Hartmann}}]{disk.photoevaporation}
{Matsuyama}, I., {Johnstone}, D., \& {Hartmann}, L. 2003, \apj, 582, 893

\bibitem[{{Mayer} {et~al.}(2002){Mayer}, {Quinn}, {Wadsley}, \&
  {Stadel}}]{2002Sci...298.1756M}
{Mayer}, L., {Quinn}, T., {Wadsley}, J., \& {Stadel}, J. 2002, Science, Volume
  298, Issue 5599, pp.~1756-1759 (2002)., 298, 1756

\bibitem[{{Murray} {et~al.}(1998){Murray}, {Hansen}, {Holman}, \&
  {Tremaine}}]{1998Sci...279...69M}
{Murray}, N., {Hansen}, B., {Holman}, M., \& {Tremaine}, S. 1998, Science, 279,
  69

\bibitem[{{Pollack} {et~al.}(1996){Pollack}, {Hubickyj}, {Bodenheimer},
  {Lissauer}, {Podolak}, \& {Greenzweig}}]{1996Icar..124...62P}
{Pollack}, J.~B., {Hubickyj}, O., {Bodenheimer}, P., {Lissauer}, J.~J.,
  {Podolak}, M., \& {Greenzweig}, Y. 1996, Icarus, 124, 62

\bibitem[{{Pringle}(1981)}]{1981ARA&A..19..137P}
{Pringle}, J.~E. 1981, \araa, 19, 137

\bibitem[{{Rasio} \& {Ford}(1996)}]{1996Sci...274..954R}
{Rasio}, F.~A., \& {Ford}, E.~B. 1996, Science, 274, 954

\bibitem[{{Shakura} \& {Sunyaev}(1973)}]{1973A&A....24..337S}
{Shakura}, N.~I., \& {Sunyaev}, R.~A. 1973, \aap, 24, 337

\bibitem[{{Shu} {et~al.}(1993){Shu}, {Johnstone}, \&
  {Hollenbach}}]{1993Icar..106...92S}
{Shu}, F.~H., {Johnstone}, D., \& {Hollenbach}, D. 1993, Icarus, 106, 92

\bibitem[{{St{\" o}rzer} \& {Hollenbach}(1999)}]{1999ApJ...515..669S}
{St{\" o}rzer}, H., \& {Hollenbach}, D. 1999, \apj, 515, 669

\bibitem[{{Tinney} {et~al.}(2002){Tinney}, {Butler}, {Marcy}, {Jones}, {Penny},
  {McCarthy}, \& {Carter}}]{2002ApJ...571..528T}
{Tinney}, C.~G., {Butler}, R.~P., {Marcy}, G.~W., {Jones}, H.~R.~A., {Penny},
  A.~J., {McCarthy}, C., \& {Carter}, B.~D. 2002, \apj, 571, 528

\bibitem[{{Valenti} {et~al.}(2000){Valenti}, {Johns-Krull}, \&
  {Linsky}}]{2000ApJS..129..399V}
{Valenti}, J.~A., {Johns-Krull}, C.~M., \& {Linsky}, J.~L. 2000, \apjs, 129,
  399

\bibitem[{{Ward}(1997)}]{1997Icar..126..261W}
{Ward}, W.~R. 1997, Icarus, 126, 261

\bibitem[{{Wilhelm} {et~al.}(1998){Wilhelm}, {Lemaire}, {Dammasch}, {Hollandt},
  {Schuehle}, {Curdt}, {Kucera}, {Hassler}, \& {Huber}}]{1998A&A...334..685W}
{Wilhelm}, K., {Lemaire}, P., {Dammasch}, I.~E., {Hollandt}, J., {Schuehle},
  U., {Curdt}, W., {Kucera}, T., {Hassler}, D.~M., \& {Huber}, M.~C.~E. 1998,
  \aap, 334, 685

\end{thebibliography}

\clearpage

\end{document}